\documentclass[twocolumn,prb,showpacs]{revtex4}

\usepackage{graphicx}
 \usepackage{rotating}
\usepackage{amsmath}
\usepackage{amsfonts}
\usepackage{amssymb}
\usepackage{enumerate}
\usepackage{longtable}
\usepackage{dcolumn}
\usepackage{bm}

\begin{document}

\title{ Non-substitutional  single-atom  defects in the
Ge$_{1-x}$Sn$_{x}$  alloy } 
\author{C.  I. Ventura$^1$,  J.D. Fuhr$^1$}
\author{R. A. Barrio$^2$} 
\affiliation{$^1$ Centro At\'omico Bariloche, 8400 - Bariloche, Argentina,
\email{ventura@cab.cnea.gov.ar    /    fuhr@cab.cnea.gov.ar}}
\affiliation{$^2$ Instituto de F{\'{\i}}sica, U.N.A.M., 01000 - Mexico
D.F., Mexico, \email{barrio@fisica.unam.mx} } 
\date{\textrm{\today}}

\begin{abstract}
Ge$_{1-x}$Sn$_{x}$ alloys have proved  difficult to form at large $x$,
contrary  to   what  happens  with  other   group  IV  semiconductor
combinations. However,  at  low  $x$   they  are  typical  examples  of
well-behaved   substitutional  compounds,   which  is   desirable  for
harnessing  the electronic properties  of narrow  band semiconductors.
In  this  paper,  we  propose   the  appearance  of  another  kind  of
single-site defect  ($\beta-Sn$), consisting of a single  $Sn$ atom in
the  center  of   a  $Ge$  divacancy,  that  may   account  for  these
facts.  Accordingly,   we  examine  the   electronic  and  structural
properties of these alloys by performing extensive numerical ab-initio
calculations  around   local  defects.  The  results   show  that  the
environment of  the $\beta$ defect relaxes towards  a cubic octahedral
configuration, facilitating the nucleation  of metallic white tin and its
segregation, as found in amorphous samples. Using the information
stemming  from these  local  defect calculations,  we  built a  simple
statistical model to investigate  at which concentration these $\beta$
defects can be formed  in thermal equilibrium.  These results agree
remarkably  well with experimental  findings, concerning  the critical
concentration above  which the homogeneous alloys cannot  be formed at
room temperature.  Our model also  predicts the observed fact  that at
lower temperature the  critical concentration increases.  We also
performed  single  site   effective-field  calculations  of  the
electronic structure, which further support our hypothesis.

\end{abstract}

 \maketitle


\section{Introduction}

\label{intro}

The search for direct energy-gap materials based on group IV 
semiconductors is not new \cite{temkin,goodman,oguz,pearsall}, but has been 
hindered for a long time by sample preparation problems. The 
importance of such compounds due to their potential 
for technological applications was recognized early. A low-energy direct gap 
would enable the use for infrared applications, long-wavelength photodetector sand emiters, 
LED and infrared laser applications, i.e. in general allow the use 
for optoelectronic devices. Combined with the use of group IV elements, 
this would be ideal for compatibility and integration with the $Si$-based technology.      

While $Ge$ is an indirect-gap semiconductor (0.41 eV gap, fundamental energy band-gap 
is 0.76 eV at the diagonal L symmetry point of the fcc diamond Brillouin zone)\cite{cohen1986},
gray tin ($\alpha-Sn$) is a semimetal with the same crystal structure (no band gap, at the zone 
center its valence and conduction bands overlap about 0.42 eV)\cite{graytin}.
It was recognized that Ge$_{1-x}$Sn$_{x}$ alloys should provide tunable gap materials, 
controlled by concentration,\cite{jenkins,baldereschi,atwater} 
and a route for obtaining   direct-gap group IV systems compatible with $Si$ integrated circuits.
The first experiments providing evidence for the indirect-to-direct gap transition 
date from 1997,\cite{atwater} while only recently crystalline samples of the quality 
and stability required by device applications were prepared.\cite{menendez2002,SSC2003,ladron}  
 In the last two years further important uses for these  materials were proposed:
they were studied \cite{atwater2006} in connection with their potential use for nanostrucured 
thermoelectric cooling devices, and the higher electron and hole  mobilities
 which might be  reached by alloying and straining.\cite{cohen2007}
Thus, they might allow to overcome current integrated circuit limitations 
to develop  higher speed $Si$ microelectronics.   

The electronic structures  of most semiconductor alloys  are smooth  functions of their composition,  
thus providing a versatile tool for  device  engineering.   Alloys of elemental semiconductors 
such as  group IV  elements  $Si$  and  $Ge$,  and alloys of III-V compounds 
as GaAs, AlAs,InAs   and InP play  key roles  in  microelectronics and  optoelectronics.  
Si$_{1-x}$Ge$_x$   is  nearly ideal as it has lattice constant 
and interband optical transition energies which are essentially linear  in x.  
However,  $Si$-based   photo-detectors  e.g.  cannot  cover the optical communication
 wavelength windows (1310-1550 nm).
Pure $Ge$ barely reaches the 1550nm range, but  it doesn't grow well  on $Si$ substrates:   
this was solved  by alloying  $Ge$  with  $Sn$.    Even  $2 \%$  $Sn$  achieves  an order of magnitude 
increase of  absorption at 1550 nm,  and the alloy grows  with device quality
on $Si$ substrates. 
\cite{menendez2002,SSC2003,APL2004,menendez2006} 

 Binary  alloys of   group  IV semiconductors  are usually  easy to
prepare  at  any concentration,  but  this is  not  the  case for  the
Ge$_{1-x}$Sn$_{x}$ alloy.  Homogeneous alloys have proved difficult to
form at large $x$.  This has considerably reduced its applicability in
the  fabrication of  electronic  devices. Other  group  IV alloys  are
formed  by  simple  substitution   of  one  component  by  the  other,
maintaining  the  overall   tetrahedral  environment  of  the  diamond
lattice. This is true even in amorphous alloys.

At the  low concentrations  at which it  has been possible  to prepare
homogeneous  Ge$_{1-x}$Sn$_{x}$  alloys,\cite{menendez2002}  it  has  been
found  that $Sn$  occupies  substitutional sites.  This  results in  a
smooth  dependence  of its  electronic  properties  as  a function  of
concentration, which is desirable to tailor key quantities as the band
gap and  the density of  states at the  Fermi level. However,  at $Sn$
concentrations  higher than  20$\%$ this  picture breaks  down because
$Sn$ exhibits  the tendency to  segregate in the cubic  $\beta$ phase,
which is metallic.  This spoils the semiconducting properties.

The main point of interest of our present work is to try to understand
theoretically  the structural  and dynamical  mechanisms  that trigger
this peculiar segregation. Some authors \cite{menendez2002} attribute this
to the large difference in size between $Ge$ and $Sn$ atoms and to the
large  lattice mismatch.  This  cannot be  the whole  picture, because
segregation has also been observed to begin at the same concentrations
in amorphous alloys \cite{barrio-am}.

Our hypothesis is that the  difficulty in the formation of homogeneous
alloys can be explained by the nucleation of the metastable $\beta-Sn$
phase (  ``white tin'',  which is  a metal and  cubic) in  the perfect
tetrahedral  lattice of  the  alloy. The  tetrahedral $\alpha$  phase,
known as ``gray tin'', is a semiconductor with a small direct gap.

We  shall investigate  the mechanisms  by which  clusters of  few $Sn$
atoms  can   acquire  a  cubic   symmetry  in  the   tetrahedral  $Ge$
matrix.  Accordingly, we  assume that  there  exists a  defect with  a
single  $Sn$ atom  bonded  to  six neighbouring  $Ge$  atoms. This  is
possible  due  to  the  important  role of  the  $d-$orbitals  in  the
electronic structure of both atoms.  The existence of such a defect in
amorphous   alloys  has   been  confirmed   by   detailed  M\"ossbauer
experiments   \cite{barrio-am}    which   show   a   signal
corresponding to a $Sn$ atom in an octahedral environment, besides the
expected signal of  the substitutional tetrahedral $\alpha-Sn$ defect. 
Signatures of the presence of two sites for the $Sn$ atoms were also found 
with $^{119}Sn$ M\"ossbauer spectroscopy in nanoscale mixtures of $Ge$/$Sn$.\cite{mrs}
 This point defect, that we shall call $\beta-Sn$  defect from now on, can  be envisaged as  a single $Sn$  atom at the center  of a
$Ge$  divacancy. This defect  (although rare)  should produce  a large
negative (shrinking)  elastic field  around it,  opposite to  the usual
positive (expansive) elastic  field of  the $\alpha-Sn$ defect,  due to
the larger size of $Sn$ as compared to $Ge$.

In the following section we will present our hypothesis for the 
incorporation of $Sn$ in the $Ge$ lattic in more detail, in Section III 
we will describe the different approaches we have used to tackle 
the problem sistematically as well of the main results obtained with each, 
providing a quite complete scenario and consistent explanation for the  
experimental findings in the framework of the hypothesis we investigated.
Finally we will present our conclusions, discussing how our results 
might be useful to solve problems for sample preparation with device quality,  
as required for the important technological applications envisaged for these 
alloys, and give an outlook to further interesting research on these materials.      

\section{Hypothesis: mechanisms for $Sn$ incorporation in the $Ge$ lattice}

\label{proposal}

The  proposed  mechanism  goes  as  follows:  small  amounts  of  $Sn$
incorporate  very easily  in isolated  places  in the  $Ge$ matrix  as
$\alpha $ defects.   The strain caused by the  size mismatch increases
with  the concentration  of  these defects  ($x_{\alpha}$), causing  a
propitious environment for the  formation of the $\beta$ defect, since
that would release the strain in the lattice.  As the $\alpha$ defects
get closer  to the  existing $\beta$ defects  they attract  each other
through  their elastic  fields,  and if  they  merge, further  elastic
energy would  be released and the  small clusters of  $Sn$ atoms would
relax to a cubic symmetry. Therefore, if difusion allows the migration
of defects during the formation,  the material would present a natural
tendency to  segregate the $Sn$ clusters  to its surface,  in order to
attain equilibrium.

All these ideas have to  be tested either by numerical calculations or
statistical models.  First of  all, one needs accurate {\it ab-initio}
electronic calculations to estimate  the scale of energies involved in
the immediate environment around  each defect. These calculations also
provide  the relaxed configurations  of the  defects, allowing  one to
test  the cubic symmetry  locally.  Furthermore,  by looking  at small
displacements within  the clusters one can estimate  the dependence of
the electronic  energy with the  volume, which is a  pressure directly
related  to  the  elastic  field  caused by  the  defects.  All  these
quantities will  allow to  construct a model  for the behavior  of the
macroscopic  system, which  will be  needed to  test if  one  can form
$\beta$ defects in  the alloy, and at which  concentrations they would
appear.

The electronic  properties of  the alloy can  be tested  by performing
effective-field calculations, like the virtual crystal approximation (VCA) \cite{VCA}  
or the coherent potential approximation (CPA).\cite{soven} It is well-known that
these  approximations  give  excellent predictions  in  substitutional
alloys, thus a failure  to describe the experimental findings suggests
a further indication of the  presence of $\beta$ defects.  In fact, we
noticed  that  if  one  takes  into  account  the  structural  changes
undergone  as a  function of  $Sn$ concentration,  one can  extend the
range of validity of these approximations.

In  the  following section  we  will  present  in detail  these  three
approaches to  the problem,  and discuss the  results in terms  of the
main hypothesis presented here.

\section{Numerical and mean field studies}

\label{approach}

\subsection{Local defect electronic calculations}

\label{Wien}

A good starting point to examine the properties of local defects in an
alloy is to perform {\it  ab initio} electronic calculations using the
common techniques of density functional theory.

We have basically two different point defects to be studied: 1) a $Sn$
atom sitting  substitutionally in a perfect $Ge$  diamond lattice, and
2) a single  $Sn$ atom  in a cage  of six  $Ge$ atoms that  surround a
double vacancy  in the perfect diamond  lattice. This cage  has a long
axis along the (111) direction, but it is not clear if the presence 
of the impurity  inside the cage should modify  this symmetry. 

The electronic energies depend on the local environment around the defect and the respective relaxation of the lattice, which involves changes  in lattice parameter (or cell volume)  
and in the individual cell-atom positions.  
We have considered unit cells of different sizes around the point defect,  in order to address these issues. 

On one hand,  we analized to what degree the energies and the lattice relaxation around the defect depend on the size of the cell, and estimated the elastic field caused by the defect. 
On the other hand, 
we  studied more complex defect environments involving several $Sn$ impurities.
We have considered unit cells of  8-atoms, as the ones studied  in previous works~\cite{atwater, ferhat}, as well as some larger cells, with 16 and 64 atoms, 
 designed to better assess the preferred configurations when multiple defects are present. 

We have determined the defect energies per atom,
 relative to the pure $Ge$ and $Sn$ cohesive energy reference values,
 allowing  {\it full  relaxation} of the different cells  studied  with respect  to the $Ge$ lattice (i.e. allowing changes of  lattice parameter   
and  of  individual positions of the atoms in the cell) . 
Specifically, for a unit cell consisting of $n$ $Ge$ atoms and $m$ $Sn$ atoms, we define the energy of the defect ($E_d$) as the cohesive energy per atom:
\begin{equation}
E_d = E_{{\rm Cell}({\rm Ge}_n{\rm Sn}_m)} - n E_{{\rm Ge-bulk}} - m E_{{\rm Sn-bulk,}}
\end{equation}
where $E_{{\rm Cell}({\rm Ge}_n{\rm Sn}_m)}$ is the total energy of the fully relaxed cell with the defect, divided by the total number of atoms in the cell.  
$E_{\rm Ge-bulk}$ and$E_{\rm Sn-bulk}$ are the bulk
$Ge$ and $Sn$ cohesive energies per atom, corresponding to perfect diamond lattices of $Ge$ and gray tin ($\alpha$-Sn).
We have also calculated the {\it partially  relaxed}  local defect energies $(E_{d}^{V_{Ge}})$ per cell atom:  
fixing the cell volume at  the value $V_{Ge}$ corresponding to bulk $Ge$,  allowing only  relaxation of the positions of the individual atoms inside the cell.

In order to estimate the elastic field caused by the presence of the
defect, we also determined the pressure exerted by each type of local
defect on the lattice. This is obtained from the dependence of the
total energy with the volume (i.e.  using  partially relaxed energies  at fixed  cell volume $E^V$). We define the pressure as:
 \begin{equation}
\label{pressure}
P = -\frac{1}{V_{Ge}}\left .\frac{\partial E^{V}}{\partial
  V}\right|_{V=V_{Ge}}
\end{equation}
calculated at the unit cell volume which corresponds to bulk $Ge$. In our convention, a positive pressure $P$
will mean that expansion of the lattice with respect to bulk $Ge$ is favoured.

The numerical calculations were done using the
full-potential linearized/augmented plane wave plus local orbital
(L/APW+lo) method, as implemented in the WIEN2K
code\cite{j1,j2,j3}. The exchange-correlation effects were treated
within the GGA (generalized gradient approximation) using the
Perdew-Burke-Ernzerhof form \cite{j4}. The radii of the muffin tin
spheres ($R_{MT}$) were chosen to be 2.25 Bohr for both $Ge$ and $Sn$ atoms. The cut-off parameter $R_{MT}$-$K_{max}$ for limiting the number of the plane waves was set equal to 8, with $K_{max}$ (= 8/2.25 =
3.55 Bohr$^{-1}$) the largest reciprocal lattice vector used in the
plane wave expansion. To integrate inside the Brillouin zone (BZ) we
used a $k$ sampling with ($7\times 7\times 7$), ($6\times 6\times 6$)
and ($4\times 4\times 4$) Monkhorst-Pack \cite{j5} meshes, for 
unit cells with 8, 16 and 64 atoms respectively. In all cases, all
atoms in the unit cell were allowed to fully relax independently and,
except otherwise stated, also lattice parameter (i.e. cell
volume) relaxation was allowed. In Table \ref{table1} we summarize the results of our {\it ab-initio} 
electronic structure calculations. The first column  identifies each of the local configurations considered. 
We have labelled the unit cells by explicitly indicating the  number of  atoms 
 of  each  species  they contain,  further details will be given below. 

\begin{table}[b!]
\caption{Results   of  electronic calculations  using Wien-2K code
($ L/APW+lo$ with $GGA$) for the cells with configurations
 described in the text. Successive columns report: corresponding ``alloy concentration'' $(x)$;
lattice parameter $(a)$ of fully relaxed crystal structure;
cohesive energy $(E_c)$ (relative to bulk $Ge$ and $Sn$);
cohesive energy per cell atom: fully relaxed  local defect energy $(E_d)$;
pressure $(P)$ due to defect; and partially relaxed local defect energy
$(E_{d}^{V_{Ge}})$ per cell atom: at $Ge$ cell volume.  All energies  given in eV.}
\label{table1}
\begin{tabular}{lcccccc}
\multicolumn{1}{l}{   Cell}  &\multicolumn{1}{c}{\quad   $x$ } &\multicolumn{1}{c}{\quad   $a$  (\AA)}&
\multicolumn{1}{c}{\, $E_c$ }
&\multicolumn{1}{c}{\, $E_d$
}  &\multicolumn{1}{c}{\,  $P$   (GPa)}
&\multicolumn{1}{c}{\,  $E_{d}^{V_{Ge}}$}
\\
\hline $Ge2$ &\quad  0  & 5.77 &  \, 0 & \, 0  & \, 0  & \, 0
\\ $Sn2$  &\quad   1 & 6.65  &
\, 0 &  \, 0  &\, - & \, -
\\  $ Ge7\,  Sn1$ &\quad   0.125 & 5.88  & \,
0.237 & \, 0.030 &\, 3.79  & \, 0.046
\\ $ Ge6 \, Sn2_{||} $ & \quad   0.25 & 6.00
& \,  0.415 & \, 0.052  &\, 7.99 & \, 0.119
\\ $  Ge15\, Sn1 $&\quad   0.062 &
5.82  &  \,   0.287  &  \,  0.018  &\,   1.81 & \, 0.022
\\  $Ge14\,Sn1_{\beta}$& \quad   0.066 &  5.72 & \, 1.804
& \, 0.120 &\, - 0.98 & \, 0.122
\\ $Ge14\, Sn2_{||} $&\quad   0.125 & 5.89 & \, 0.652 & \, 0.041 &\,
3.99 & \, 0.059
\\ $Ge14\, Sn2 $ &\quad   0.125 & 5.90 & \, 0.462 & \, 0.029 &\,
3.86 & \, 0.046
\\  $ Ge11\,  Sn5 $ &\quad   0.31 &  6.10 &  \, 1.221 &  \, 0.076
&\, 11.01 & \, 0.193
\\ $ Ge11\, Sn4_{a} $ &\quad   0.26  & 5.85& \, 1.330 & \,
0.089 &\, 1.85 & \, 0.095
\\ $ Ge11\, Sn4_{b}  $ &\quad   0.26 & 5.84 & \, 1.333 &
\, 0.089 &\, 1.85 & \, 0.095 \\
$Ge64$ &\quad   0 & 5.76 &  \, 0 & \, 0  & \, 0 & \, 0
\\ $Sn64$  &\quad   1 &  6.63  &
\, 0 &  \, 0  &\, - & \, -
\\  $ Ge56\, Sn6$ &\quad   0.097 & 5.77 & \, 2.624
&  \, 0.042  &\, 0.47 &\, 0.112
\\  $ Ge56\,  Sn7_{\beta'} $  &  \quad    0.111 & 5.81 &
\,  1.860 &  \, 0.030  &  \, 1.71  &\, 0.033
\\  $Ge\, Sn  $ &  \quad   0.5 & 6.21 &\, 0.052 & \, 0.026 & \, 17.93
& \, 0.297 \\

\end{tabular}
\end{table}

First, we have considered the two limiting cases  of the alloy, namely the  perfect diamond lattices
of $Ge$ and gray tin, in order to extract their cohesive energies as reference values for  the local defects. 
In these cases (labelled as $Ge2$ and $Sn2$ in Table I), we used  the Wigner-Seitz  cells for  diamond, 
containing  two  atoms and periodic boundary conditions. 
We also calculated the corresponding bulk energies and lattice parameters  for cells 
including 64 atoms: $Ge64$ and $Sn64$, resulting in cohesive energies differing from the 2-atom cell values by less than 
0.1 meV.  Observe in the table, that also the lattice parameter changes are negligible.

In the 8-atom cells we studied a single  substitutional $\alpha-Sn$ defect
(the $Ge7Sn1$ case  of Table  I), and the pair defect consisting of 
 two  contiguous $\alpha$  defects in
which the two $Sn$ atoms  align parallell to the $(111)$ symmetry axis
of the  diamond lattice (case $Ge6Sn2_{||}$). In the latter case,  we have verified that starting from different
orientations of the  pair of $Sn$ atoms,   this symmetry 
is the one  corresponding to  the minimum  of the  total  relaxation process.
These 8-atom configurations were studied before, with local density approximation for 
density functional theory(LDA).\cite{atwater}
In agreement with the well-known general trends,   the
GGA yields slightly lower cohesion energies and larger lattice parameters
than  LDA,  in the cases where we can compare our results
with previous LDA calculations.\cite{atwater,baldereschi}

\begin{figure}[th]

\includegraphics[angle=0 , width=\columnwidth]{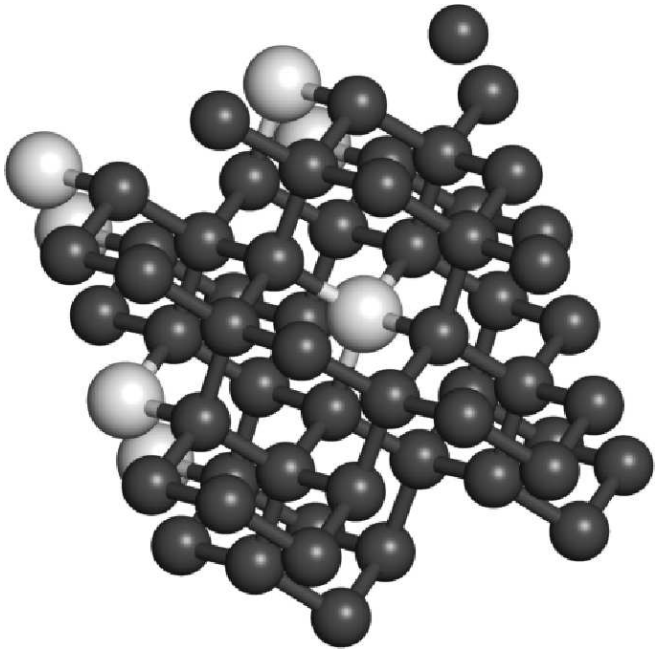}
\caption{$\alpha-Sn$ (substitutional) defects in $Ge$ lattice. ($Ge15 Sn1$)}
\label{alpha-tin}
\end{figure}

\begin{figure}[ht]
\includegraphics[angle=0 , width=\columnwidth]{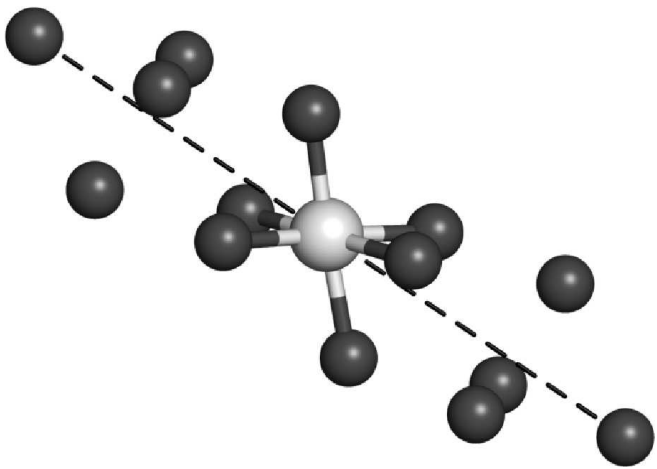}
\caption{$\beta-Sn$ defect in $Ge$ lattice: 16-site supercell shown. ($Ge14 Sn1_{\beta}$) }
\label{beta-tin}
\end{figure}

\begin{figure}[h!]
\includegraphics[angle=0 , width=\columnwidth]{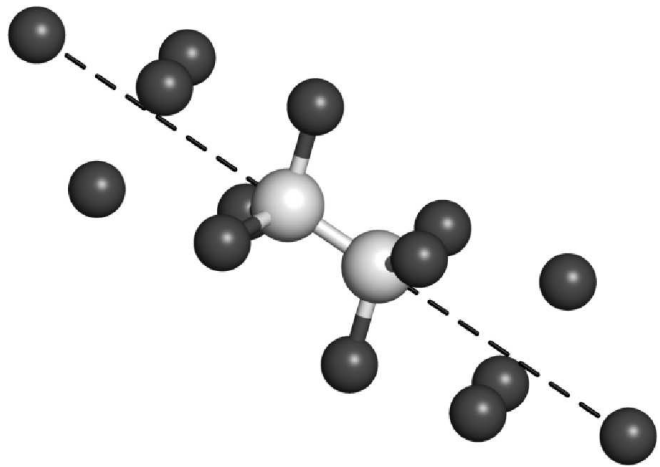}
\caption{$Sn$ pair-defect in $Ge$ lattice: 16-site supercell shown. ($Ge14 Sn2_{||}$)}
\label{pair-defect}
\end{figure}

We also studied the single  $\alpha-Sn$ defect  and the pair defect configurations in a 16-atom cell,
to see the  effect of the  size of the cell  on the relaxed  configurations. 
In particular,  the fact that the energy obtained for  
two independent (non-contiguous) $\alpha-Sn$ defects 
in a 16-site cell ($Ge14 Sn2$)  is approximately twice the energy of one $\alpha$ defect in the 8-site cell 
($Ge7 Sn1$)  is a further  indication that our results are reasonably size-independent. 

In Figure~\ref{alpha-tin} we show the 
   the substitutional $\alpha-Sn$
defect $Ge15Sn1$, where a $Sn$ atom at the  center of a 16-site cell. In Fig.~\ref{beta-tin}, we depict 
the non-substitutional $\beta-Sn$ defect with a $Sn$ atom at the center 
of a divacancy in a 16-site cell of the $Ge$ lattice (denoted  
$Ge14 Sn1_{\beta}$ case), 
while $Ge14Sn2_{||}$ labels the above 
mentioned pair defect,  now placed in a 16-site $Ge$ cell (see Fig.~\ref{pair-defect}). All figures included in this section,  
exhibit the atoms in the fully relaxed configuration obtained for each cell.

\begin{figure}[ht]
\includegraphics[angle=0 , width=\columnwidth]{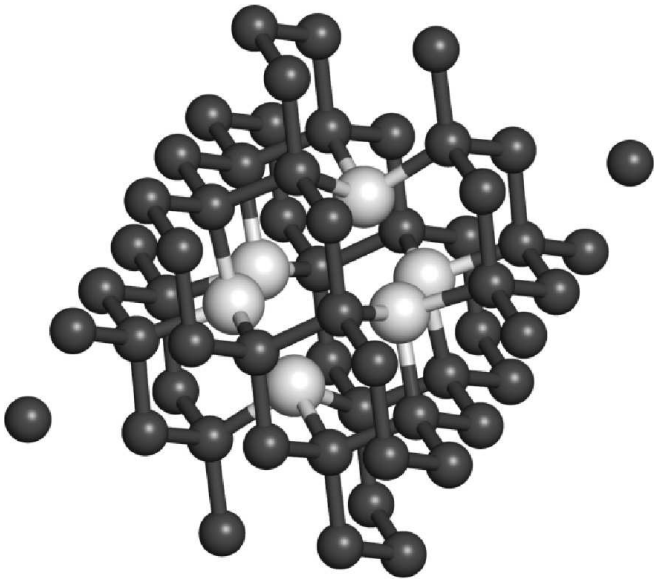}
\caption{6-$Sn$ cluster-defect with a central divacancy in $Ge$ lattice: 
64-site supercell shown.  ($ Ge56 Sn6 $)}
\label{6cluster-defect}
\end{figure}

\begin{figure}[h]
\includegraphics[angle=0 , width=\columnwidth]{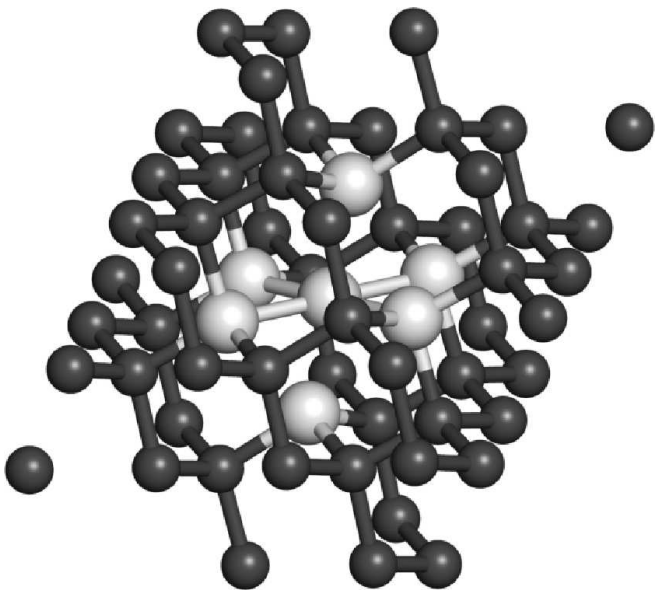}
\caption{7-$Sn$ cluster-defect ($\beta$ defect surrounded by 6 $Sn$-atoms) 
in $Ge$ lattice: 64-site supercell shown. ($ Ge56 Sn7_{\beta'} $) }
\label{7cluster-defect}
\end{figure}

For 16-site cells we have also studied the effect of agglomeration of $Sn$:  $Ge11 Sn5$ corresponds 
to five substitutional $Sn$ atoms at the  center of a 16-site cell; $Ge11 Sn4_a$ denotes the previous 
configuration with the central $Sn$ atom removed, such that a vacancy surrounded by four $Sn$ atoms 
occupies the center of the 16-site $Ge$-lattice cell,
while $Ge11 Sn4_b$ refers to the same cell containing in its center four $Sn$ atoms plus a vacancy now being placed at one of the nearest neighbour sites of the cell center. 
We have also studied two 64-site cells, shown in Figs.~\ref{6cluster-defect}
and \ref{7cluster-defect}, containing respectively a  
    6-$Sn$ cluster-defect with a central divacancy ($Ge56 Sn6$),  and 
a $\beta$ defect surrounded by 6 $Sn$-atoms ($Ge56 Sn7_{\beta'}$), always in the $Ge$ lattice.

 The last row of Table I  refers to  $GeSn$, the  x=0.5  zincblende structure, which should be strain-free. 
Notice that the pressure shown in last row of Table I, was calculated relative to the bulk-$Ge$  cell volume (Eq. ~\ref{pressure}),
instead of relative to the zincblende cell volume. 

\setlength{\unitlength}{1243sp}
\begingroup\makeatletter\ifx\SetFigFont\undefined
\gdef\SetFigFont#1#2#3#4#5{
  \reset@font\fontsize{#1}{#2pt}
  \fontfamily{#3}\fontseries{#4}\fontshape{#5}
  \selectfont}
\fi\endgroup

Our  ab-initio electronic calculations, performed in supercells larger than in previous studies\cite{atwater,baldereschi,ferhat}   
and with various  different local defects, among other facts reveals that:

i) cubic octahedral symmetry  is favoured in clusters with few $Sn$ atoms, as can be seen in the fully relaxed configurations of Figs.~\ref{beta-tin}, \ref{6cluster-defect}, and \ref{7cluster-defect}.

ii) electronic energy per atom is not much increased by accumulating  $Sn$ clusters;

iii) substitutional $\alpha-Sn$  and non-substitutional $\beta-Sn$ single-atom defects
do create opposite elastic fields:  this leads to an effective attraction between them,
 such that  if they do merge elastic energy is released  and the small cluster 
 relaxes to cubic symmetry;

iv) building up a local pressure in the lattice is not necessarily correlated
with the electronic energy of the cluster. This agrees with former ideas about
the existence of two independent difficulties for the  formation of  homogeneous alloys, namely, the large size difference between $Ge$ and Sn atoms with a 15$\%$ lattice mismatch, 
and the  d-bands  which  make gray tin unstable at room temperature;~\cite{menendez2002,ladron} 

v)  the energy of two $\alpha-Sn$ defects increases when they are closer: 
indicating it would be difficult to form a homogeneous high concentration totally substitutional alloy 
(due to the electronic energy cost).  If, as generally mentioned before, strain was
 the only/main factor jeopardizing formation: x=0.5 zincblende  ($GeSn$) should be easy to form, 
which is not the case. 

\subsection{Statistical mean field approach}
\label{critical}

We used the valuable information from our electronic calculations
of the previous  section,  to try  to
understand and predict the behavior of  the alloy as a function of the
$Sn$  concentration  (x).  Taking  advantage  of the  fact  that  both
experiments  and  calculations  indicate a  favourable  substitutional
incorporation of  $Sn$ in the  $Ge$ matrix for low  concentrations, we
could model  the statistical behavior of the  mixture during annealing
to attain thermodynamical equilibrium by a stochastic approach. At low
$x$, defects  containing only  one $Sn$ atom  are relevant.  Imagine a
tetrahedral cage  of $Ge$ atoms in which  the central four-coordinated
site could be occupied by either $Ge$ or $Sn$, the former representing
the  perfect $Ge$ lattice  ($T_{Ge}$) while  the latter  represents an
$\alpha-Sn$ defect  in $Ge$  ($T_{\alpha}$).  By substitution  of only
one $Sn$ atom in any of the corners of the cage, one gets the $6-$fold
coordinated $Sn-$pair defect studied in previous section.This $6-$fold
cage, which is the environment  of a divacancy, could also be occupied
by a  single central  $Sn$ atom, giving  the non-substitutional 
$\beta-Sn$ defect studied in previous section ($T_{\beta}$).  
The $\alpha-Sn$ defect can be transformed by substitution of a single atom into the 
pair-defect, which becomes important only for larger $x$. To keep the treatment to first order in $x$, we approximate the pair defect by two separate $\alpha-Sn$ defects, based on the fact that the numerical calculations of  previous  section  indicate that  the  energy of the pair-defect
$E_{\parallel}$   is quite similar to twice the energy of  one $\alpha-Sn$ defect, $E_{\alpha}$.
However, this energy difference will be taken into account below, when needed in the treatment.

To first order in $x$, there are only these three local configurations:
 $T_{i} (i=\alpha, \beta, Ge)$, which form a complete closed set, in
the sense that one can  transform one into the other without leaving
the  set  by  substitution of  a  single  atom.   This allows  one  to
formulate  a  dynamical process,  representing  the  formation of  the
infinite  solid by  a large  number of  such  one-step transformations
encoded  in a  stochastic matrix  (which we  will denote  by $\bf{S}$)
.\cite{paperRaf1-stochastic, book-Kerner} The elements of $\bf{S}$
can  be  considered  as   the  probabilities  $p_{a,b}$  of  obtaining
configuration $b$  starting from  configuration $a$, and  must contain
all the physical information available. Concretely, writing a one-step
transformation as:
\begin{equation} \mathbf{S} \vec{v}_0 = \vec{v}_1 ,\end{equation} 
where $\vec{v}_0$ is a vector whose three components represent 
the  initial  concentrations  of  each  of  the  three  configurations
$T_{i}$,  therefore  the  sum  of  its  three  components  equals  one
(i.e.  $norm_1  (\vec{v}_0)  =  1$).   In general,  the  final  vector
$\vec{v}_1$ is different but its  $norm_1$ should be conserved.
This is guaranteed by normalizing the columns of matrix $\bf{S}$ to one,  so that  $\bf{S}$  becomes a
stochastic matrix. A  well-known property of stochastic matrices  is that one
of their eigenvalues $\lambda_1$ is equal to $1$ while all others have
magnitudes which are  less than $1$ ($  | \lambda_i | < 1  $) .  After
repeating  the  transformation  $n$  times, the  transformed  vector
can be  expressed in terms  of the eigenvalues and
eigenvectors $\hat{e}_i$ of the stochastic matrix $ \bf{S}$ as,
\begin{equation}  
\mathbf{  S  }^{n}  \vec{v}_0  =  \sum_{i=1}^n  \lambda_i^{n}  a_i
\hat{e}_i ,
\end{equation}  where $a_i$  are the components  of $\vec{v}_0$  in the
eigenvector basis.  If $n \longrightarrow \infty$,  the only surviving
element of the summation above is:
\begin{equation}    \mathbf{S}^{n}  \vec{v}_0   \sim  (\lambda_1)^n  a_1  \hat{e}_1  =
\lambda_1 a_1 \hat{e}_1 = a_1  \hat{e}_1 .\end{equation}
  Taking into account that
$norm_1 (\vec{v}_0) = 1 =  norm_1 (\hat{e}_1)$, and that $\bf{S}$ is a
stochastic  transformation,  we  immediately  obtain:  $ a_1  =  1  $.
Therefore, the  elements of $  \hat{e}_1$ should be considered  as the
concentrations  of   the  final  configurations   in  the  macroscopic
system.  For  any  arbitrary  initial  concentrations,  i.e.  for  any
$\vec{v}_0$, under  homogeneous formation conditions,  one attains the
fixed point:
\begin{equation}
 \hat{e}_1 = (x-y, y, 1-x),
\label{eigen}
\end{equation}
where $y$ is the concentration of $\beta-$defects in the sample (and
$x-y$ denotes the concentration of $\alpha-$defects).

The transition probabilities  $p_{a,b} = P_{(a,b)}/N_{a,b}$ ($N_{a,b}$
are the required normalization factors) are composed of three parts. A
first factor is the conditional probability of obtaining configuration
$b$,  starting from  configuration $a$,  that is:  $x_{a}  x_{b}$. The
second factor is a number  counting all the possibilities of arranging
the chemical bonds without changing the final results of the transition, and it
could   be   interpreted   as   a  sort   of   local   configurational
entropy. Finally, the energy barriers to perform the transition should
be considered. If the time scale  for a chemical bond to reach thermal
equilibrium is much smaller than the time it takes for the whole solid
to  equilibrate, then the  third part  can be  written as  a Boltzmann
factor containing  the temperature of the  bath. In our  case, we thus
have: 
 \begin{eqnarray}  P_{\alpha-Ge} &=&  (1-x)(x-y)  4!  {\rm  e}^{
-E_{\alpha} / (k_B T)}; \nonumber  \\ 
P_{\beta-Ge} &=& (1-x) y 6!  {\rm
e}^{ -E_{\beta}  / (k_B T) };  \nonumber \\ 
P_{Ge-Ge}  &=& (1-x)^{2} 4!
{\rm e}^{ -E_{Ge} / (k_B T)} .
\label{Pes}
\end{eqnarray}

We shall  consider the remaining transition probabilities  to be zero,
since they are second order  in $y$, except for the transition between
two    neighbouring     $\alpha$    configurations    described    by:
\begin{equation} P_{\alpha-\alpha}  = \, (x-y)^2  \, \, 4^2 3^2  \, \,
{\rm e}^{ -(E_{\|} -E_{\alpha})/ (k_B T),} \label{Paux}
\end{equation} 
due to  the energy considerations which were  mentioned above. Observe
 that  the energy  barrier is  only  the difference  between the  pair
 defect and the $\alpha$-defect, since only one step is needed to form
 it from the initial $\alpha$ configuration.

Summarizing, the stochastic matrix  $\bf{S}$ takes the following form:
  \begin{equation} \mathbf{S} = \left (
\begin{array}{ccc}
A & 0 & B \\ 0 & 0 & C \\ (1-A) & 1 & (1-B-C)
\end{array}
\right ), \label{Smatrix}
\end{equation} 
where:   
 \begin{eqnarray}  
  A    &=&   \frac{    P_{\alpha-\alpha}}{
 P_{\alpha-\alpha}  +   P_{\alpha-Ge}};  \nonumber  \\   B  &=&  \frac{
 P_{\alpha-Ge}}{ P_{\alpha-Ge} +  P_{\beta-Ge} + P_{Ge-Ge} }; \nonumber \\ 
 C  &=&  \frac{  P_{\beta-Ge}}{  P_{\alpha-Ge}  +  P_{\beta-Ge}  + P_{Ge-Ge} } .
\label{ABC}
\end{eqnarray}

Notice  that,   due  to   the  normalization  condition,   only  three
independent energies  appear in  the problem, as  can be  verified by
respectively dividing  numerator and denominator  of $A,B$ and  $C$ by
$P_{Ge-Ge}$.  Thus,  only the three  relative energies of  each defect
configuration  with  respect  to  the $Ge$  lattice  ($\tilde{E}_{\mu}
\equiv E_{\mu}-E_{Ge}$) will appear,  which are precisely the energies
of the local defects calculated numerically in the previous section.

The eigenvector of $\bf{S}$ with eigenvalue $1$ is:
\begin{equation} \hat{e}_1 = ( x_{\alpha}, x{_\beta}, x_{Ge}) ,\end{equation}
 where:
\begin{eqnarray}
 x_{\alpha} &=& \frac{ B} { 1-A  } x_{Ge} ;\nonumber \\ x_{\beta} &=& C
 x_{Ge}; \nonumber \\ x_{Ge} &=& \frac{ 1 - A} { (1-A)(1+C) + B } ,
\label{xes}
\end{eqnarray}
which  due  to  the  homogeneity  condition  should  be  identical  to
Eq.~\ref{eigen}.

Therefore, at   a  fixed   temperature   $(T)$  one   can  obtain   the
concentration  of  $\beta-Sn$ defects  by  finding  the  zeros of  the
function

\begin{equation}
f(x,y) = \frac{1-A}{(1-A)(1+C)+B} -(1-x),
\label{zeros}
\end{equation}
which we do  numerically. The needed values of  the energies are taken
from Table~\ref{table1} and  are:  $\tilde{E}_{\alpha}=0.018$  eV,
$\tilde{E}_{\beta}=0.120$ eV, and $\tilde{E}_{\|}=0.041$ eV.

\begin{figure}[h!]
\includegraphics[angle=0 , width=\columnwidth]{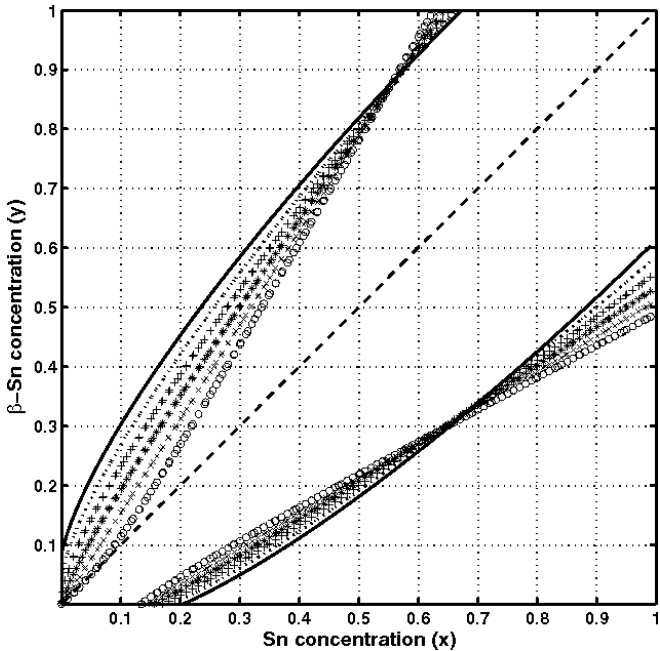}

\caption{  Concentration  of $\beta-Sn$  defects  as  a function  $Sn$
concentration in a  Ge$_{1-x}$Sn$_{x}$ alloy, at different temperatures.
Full line: $T$ =  16$^\circ$C; 39$^\circ$C (dotted); 62$^\circ$C (plus
signs), 85$^\circ$C  (asterisks), 108$^\circ$C (crosses); 131$^\circ$C
(circles). }
\label{criticallast}
\end{figure}

In Fig.~\ref{criticallast}  we depict  the zeros of  $f(x,y)$ obtained
for  six  different temperatures.   Notice  that  the only  physically
meaningful solutions are the  ones below the  $y=x$ line  (since the
$\beta-$defect  concentration  $y$ cannot  be  larger  than the  total
$Sn-$concentration  present, $x$).  Observe  that at  room temperature
there are  no $\beta$ defects present  for concentrations $  x < 0.2$,
meaning that  the alloy is perfectly  substitutional (only $\alpha-Sn$
defects  present). Therefore,  a  continuous shift  of the  electronic
properties is  to be  expected and could  be accurately modelled  by a
simple   VCA   approximation,  as   previously   noted  and   measured
\cite{jenkins,menendez2002}  and  we  will  further  discuss  in  next
section.  For larger  $Sn$-concentration, $x  > 0.2$,  one finds  that a
non-zero  concentration  of  $\beta-Sn$  defects appear,  which  would
introduce  electronic traps  in the  gap  of the  alloy. A  monotonous
increase of  the number  of $\beta$ defects  follows, as more  $Sn$ is
incorporated to the $Ge$ lattice.

Furthermore,  our simple  model predicts  that as  the  temperature of
formation  of the  solid is  increased  the $\beta$  defects start  to
appear at lower $Sn$  concentrations, as expected because the increase
in thermal energy favours the overcoming of all energy barriers.  This
fact provides an explanation  for the findings of $Ge_{1-x}Sn_x$ alloy
preparation by  chemical vapor deposition (CVD) \cite{menendez2002}, where  
 it has been  observed that
 more  substitutional  $Sn$  could  be incorporated  if  the  substrate
temperature is  lowered.  This  is also known  to happen  in amorphous
Ge$_{1-x}$Sn$_x$ alloys \cite{barrio-am}.

\begin{figure}[ht!]
\includegraphics[angle=0 , width=\columnwidth]{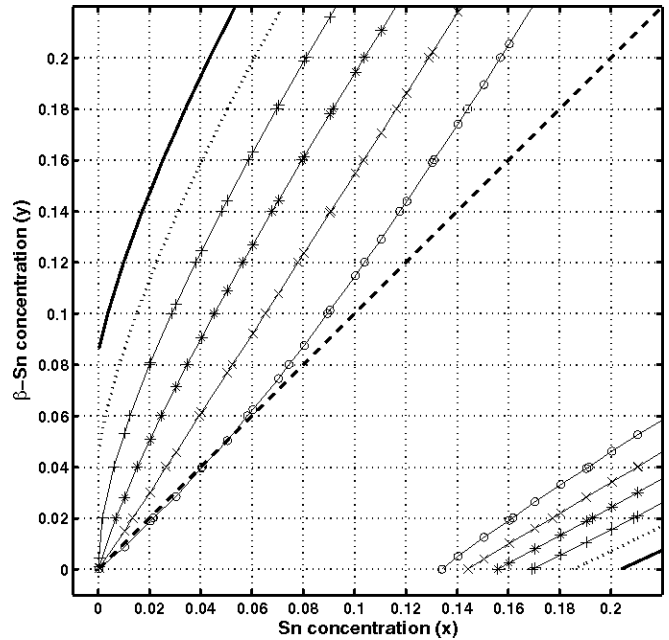}

\caption{Amplification   of   the    low   concentration   region   of
Fig.~\ref{criticallast}.   Full line:  $T$ =  16$^\circ$C; 39$^\circ$C
(dotted);   62$^\circ$C   (plus   signs),   85$^\circ$C   (asterisks),
108$^\circ$C (crosses); 131$^\circ$C (circles). }
\label{insetlast}
\end{figure}

\begin{figure}[h!]
\includegraphics[angle=0 , width=\columnwidth]{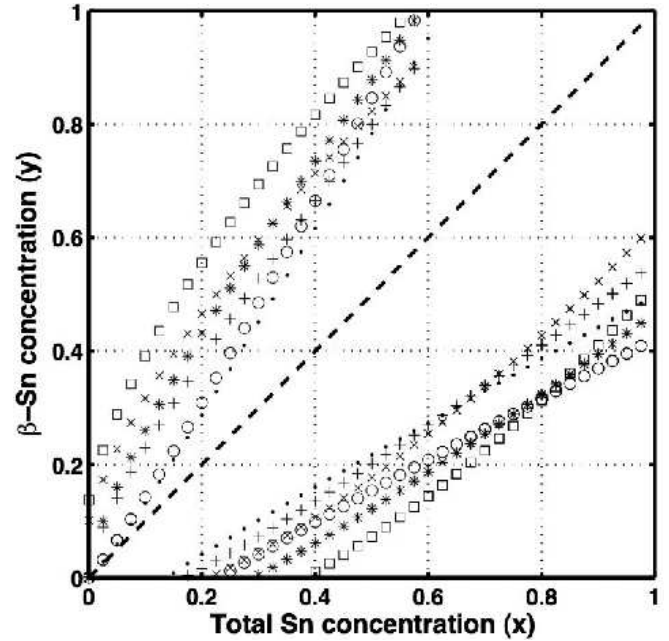}

\caption{ Comparison of use of alternative defect energy definitions: relaxed vs. fixed at $V_{Ge}$
cell volumes, on the  concentration  of $\beta-Sn$  defects  as  a function of total $Sn$ concentration.
Using $E_{d}$, at : $T$ =  18$^\circ$C (squares); 79$^\circ$C (asterisks); 134$^\circ$C (circles).
Using $E_d^{V_{Ge}}$, at the same temperatures:
18$^\circ$C  (crosses), 79$^\circ$C (plus signs); 134$^\circ$C (dots). }
\label{comparison-edef}
\end{figure}
In  Fig.~\ref{insetlast}   we  show   an  amplification  of   the  low
concentration  results,  where   the  above  discussed  dependence  on
temperature  is  clearly  seen.  Also,  notice that  for  high  enough
temperature  the non-physical  roots start  to cross  the  $x=y$ line,
meaning that $\beta$ defects  can be formed at any $Sn-$concentration.
This  is expected  to happen,  because disposing  of large  amounts of
thermal  energy  allows  the  significant  creation  of  more  complex
defects, than those considered  in this simple approximation. We think
that there  is not much  to gain by  increasing the complexity  of the
present model, since the larger  the defect space considered, the less
reliable are the local energies calculated.

Finally, in  Fig.~\ref{comparison-edef}   we show  the 
effect  of using the alternative definition for the defect energy
obtained by partial relaxation, fixing the cell volume at the bulk-$Ge$ value ($E_d^{V_{Ge}}$). 
As expected from the differences in respective energies stated in Table I, which show a 
relatively lower increase for the $\beta$-defect energy than for the $\alpha$ or pair-defect cases,
fixing the cell volume  leads to $\beta-$defects appearing at slightly lower $Sn$ concentration values.
The comparison of results  using the  alternative defect energy definitions
reveals that the main effect of using the partial relaxation is a general shift (reduction) 
of  the ``effective'' energy barriers in the Boltzmann factors of Eqs.~\ref{Pes} and ~\ref{Paux}: 
thus being analogous to a rescaling of the temperature.

\subsection{Electronic structure studies}

\subsubsection{Supercell calculations}
\label{WienDOS}

\begin{figure}[h]
\includegraphics[angle=0 , width=\columnwidth]{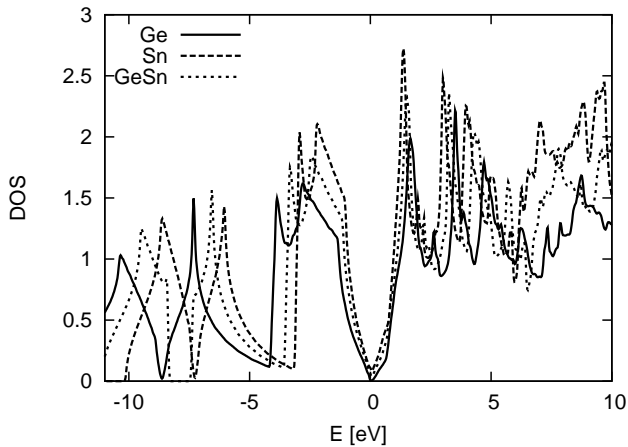}
\caption{ab-initio  (Wien-2K) calculated DOS with  16-atom supercells.}
\label{WDOS}
\end{figure}

The   {\it   ab-initio}  local   defect   calculations  presented   in
section~\ref{Wien}, also yield the  density of states and energy bands
of perfectly ordered periodic  crystals formed by those supercells. As
an illustration  we show in Figure~\ref{WDOS} the densities
of states (DOS) of three characteristic crystals, namely the pure $Ge$
and $Sn$ ones, and  the zinc-blende-type $GeSn$, previously studied in
Ref.~\cite{cardona}.

\begin{figure}[h]
\includegraphics[angle=0 , width=\columnwidth]{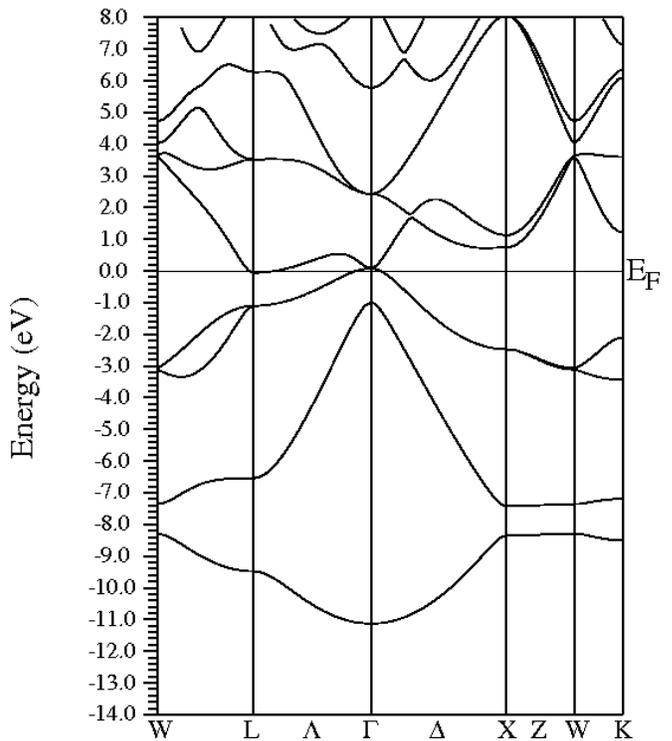}
\caption{ab-initio  (Wien-2K) calculated bands  with 2-atom  cells for
GeSn.}
\label{WGeSnband}
\end{figure}

It can be noticed in  Figure~\ref{WDOS} that the $Ge$ bands 
are slightly wider than the $Sn$ ones, and that the density of states 
of $GeSn$ at the Fermi level lies 
between the  higher value for $Sn$  and the very  small value obtained
for $Ge$, as one would expect.  Notice that the pure $Ge$ DOS does not
have  a gap  at  the  Fermi level,  which  is not  a  surprise due  to
inaccuracies at small energies around  the Fermi level of the approach
GGA-Wien2K.  Another peculiar  thing is the fact that  deep down the valence
band  (around -8eV),  both  pure  systems present  a  single Van  Hove
singularity with zero DOS corresponding  to degeneracy of the bands at
the  $X$  point  of  the  first  Brillouin zone  of  the  fcc  diamond
structure. This degeneracy is due  to the inversion symmetry, which is
lacking in the  zincblende structure.  The wide gap  in the density of
states around those energies exhibited in zinc-blende $GeSn$ is due to
the removal of this degeneracy.\cite{cardona} Of course, such features
due  to symmetry-breaking  are not  expected in  the  real crystalline
alloys, since perfect order is not realized.
All this can be seen more clearly in Figure~\ref{WGeSnband}, where the
electronic  bands from  the 2-atom  cell  are shown.  Observe that the  lowest conduction band  crosses  the Fermi level at point $L$.

\subsubsection{ TB + VCA}
\label{tb+vca}

Tight-binding  (TB) calculations  for electronic  band  structures are
very  useful because,  besides  their simplicity,  it  is possible  to
incorporate    many   features   by    suitably   choosing    the   TB
parameters.   These  method  applied   to  $Ge$   proved  not   to  be
straightforward,  since  a   naive  calculation  considering  $sp^{3}$
orbitals is  unable to reproduce essential features  like the indirect
band  gap  and the  bandwidths.  This  is due  to  the  mixing of  $d$
electrons in the  conduction bands of $Ge$, which for  $Si$ or $C$ are
not  important, while  for $Sn$  and  $Pb$ are  responsible for  their
metallic  behavior. The  role of  $d$ electrons  could be  mimicked by
introducing a  pseudo-orbital ($s^{*}$)  without increasing the  size of
the TB Hamiltonian.  Finer properties, such as the exciton spectrum of
$Ge$ cannot be explained without introducing spin-orbit interactions.

\begin{figure}[ht!]
\includegraphics[angle=270 , width=\columnwidth]{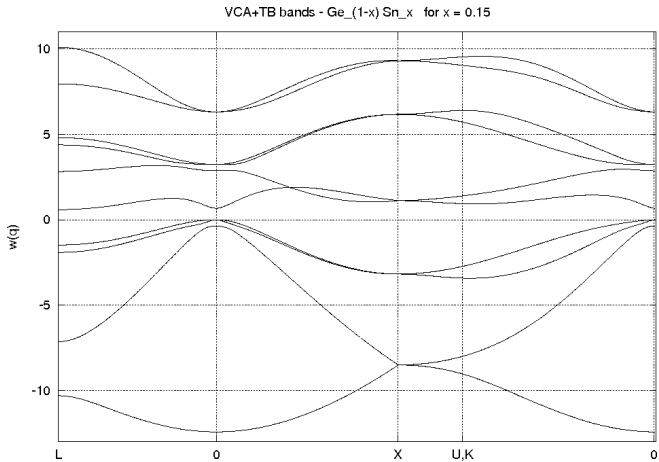}
\caption{TB+VCA   calculated    bands   as  in  Ref. ~\cite{jenkins}   for
Ge$_{1-x}$Sn$_{x}$ at x=0.15.}
\label{TBVCA015band}
\end{figure}

In  1987, Jenkins  and Dow  \cite{jenkins} presented  a  complete TB
model for  the Ge$_{1-x}$Sn$_{x}$ alloy,  including spin-orbit, $s^{*}$,
and second-neighbour  interactions.  They examined several properties
by using a simple virtual crystal approximation (VCA) \cite{VCA} description 
of the alloy. Their results  are in very  good  agreement  with  experiments 
for  low  concentrations,  in particular they predict the indirect to direct gap transition to occur
at $x \approx  0.15$.  For the sake of  completeness and to facilitate
comparison with the low concentration regime, we have reproduced their
TB+VCA  calculations and  in Figure~\ref{TBVCA015band}  we  depict the
electronic bands precisely at this critical concentration. Notice that
the optical  gap is the same at  the center of the  1st Brillouin zone
and at point $L$.\cite{jenkins}

\begin{figure}[ht]
\includegraphics[angle=270 , width=\columnwidth]{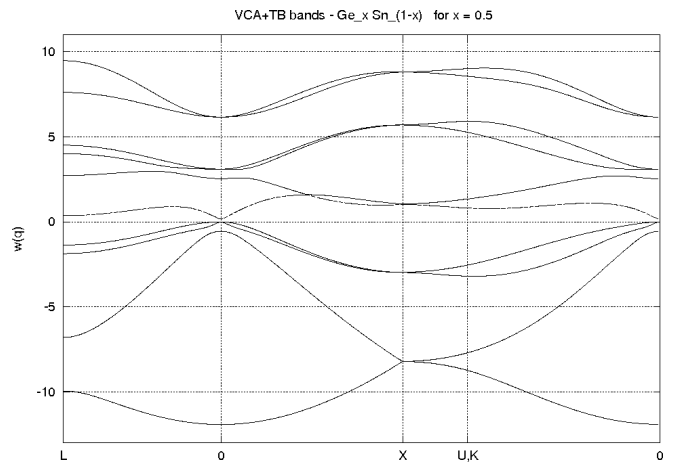}
\caption{TB+VCA   calculated    bands   as  in  Ref. ~\cite{jenkins} for
Ge$_{1-x}$Sn$_{x}$ at x=0.5.}
\label{TBVCA05band}
\end{figure}

The dependence of electronic properties as a function of concentration
continues smoothly in this approximation, contrary to the experimental
facts. To illustrate this, in Figure~\ref{TBVCA05band} we plot the bands
at 50$\%$ concentration,  where a small energy gap  at the Fermi level
is  still  present.  Spectroscopic ellipsometry  and  photoreflectance
experiments  \cite{menendez2006}  found  that  there is  a  strong  bowing
tendency,  i.e. non-linear  behavior of  this gap  with concentration,
that would close the gap at  a much lower value: $x \approx 0.37$.  As
one  expects for the  alloy, no  degeneracy is  removed at  point $X$,
because  it does  not introduce  spurious order  like  the zinc-blende
bandstructure of Fig.~\ref{WGeSnband}.

In  summary, these  TB+VCA  calculations  can be  trusted  only at  low
concentrations,  where  they  provide  a  valuable tool  to  interpret experiments.

\subsubsection{Coherent Potential Approximation}
\label{cpa}

In many random alloys, the coherent potential approximation (CPA) \cite{soven} for the
treatment of substitutional disorder leads to an improved description of the electronic
structure, overcoming some limitations of the VCA discussed in previous section.  
CPA has been applied successfully to many semiconductors,\cite{cpa-semic} and
in particular to Ge$_{1-x}$Si$_{x}$ alloys \cite{cpa-gesi} where it
predicts moderate bowing in the optical transition energies. Concerning 
Ge$_{1-x}$Sn$_{x}$, special supercell calculations have been performed to mimic the alloy 
by averaging over selected structures \cite{ferhat,harrison}. They  predict a large and 
compositional-dependent direct gap bowing, in agreement with experimental reports.\cite{menendez2006}
 
CPA electronic structure studies of the Ge$_{1-x}$Sn$_{x}$ alloy were
lacking, up to now. This has prompted us  to undertake them,  
as a further way  to indirectly test our hypothesis  
regarding the relevance of non-substitutional defects  in this alloy. 
In general,
effective-field calculations, like VCA  or CPA, give  excellent predictions 
 of  the electronic properties  in  substitutional alloys, thus a failure  
to describe the experimental findings suggests a further indication 
of the  presence of $\beta-Sn$ defects.  
In fact, we found that  if  one  takes  into  account  the  structural  changes
undergone as a  function of  $Sn$ concentration,  
one can extend the range of validity of these approximations.
   
\begin{figure}[ht!]
\includegraphics[angle=270, width=\columnwidth]{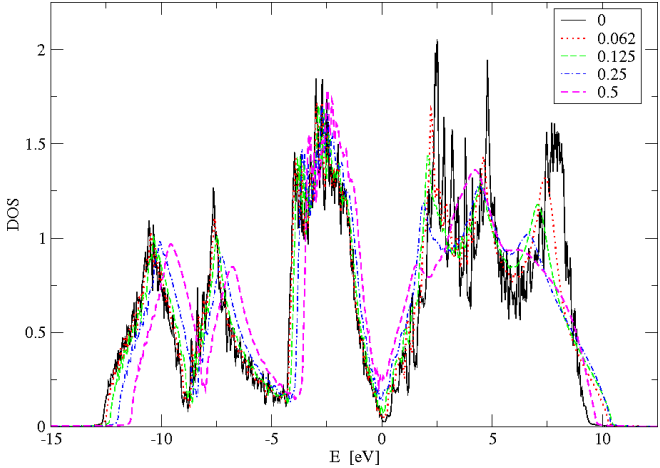}
\caption{(Color online) CPA for Ge$_{1-x}$Sn$_{x}$, via FPLO-5 code:  total density of states
  for the $\alpha-Sn$ concentrations  indicated inside plot.  
  Lattice parameters  from respective Wien-2K  obtained structural data.}  
\label{CPA-DOS-0to05}
\end{figure}

\begin{figure}[ht!]
\includegraphics[angle=270, width=\columnwidth]{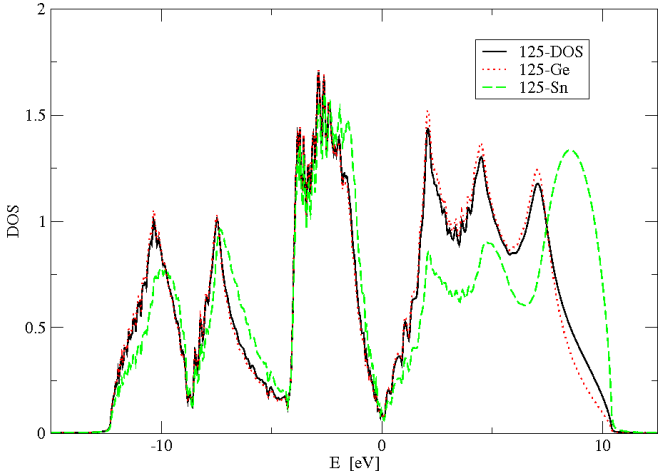}
\caption{(Color online) CPA results for Ge$_{0.875}$Sn$_{0.125}$, via FPLO-5:  
total and partial (species-resolved) DOS for each alloy component, as a function of energy. }
\label{CPA-DOS-125}
\end{figure}

In the following, we report results of our CPA calculations for the
Ge$_{1-x}$Sn$_{x}$ alloy, using the FPLO-5 code developed by the IFW-Dresden
group.\cite{FPLO,FPLO-CPA} 
The FPLO-5 package is a full-potential local-orbital minimum-basis code,
combining {\it ab-initio} LSDA (local spin density approximation)
bandstructure treatment\cite{FPLO} with single-site CPA routines  
for substitutional disorder\cite{FPLO-CPA}  (it employs the Blackmann-Esterling-Berk 
single site-CPA extension for combined diagonal and non-diagonal disorder \cite{BEB}).

In Fig.~\ref{CPA-DOS-0to05} we show the FPLO+CPA total density of states for
Ge$_{1-x}$Sn$_{x}$ as a function of energy,  at different substitutional-$Sn$ concentrations:
$x= 0.0, 0.062, 0.125, 0.25, 0.5$. Notice that a smooth behavior
as a function of concentration is obtained, with changes of the bandwidth,
 and a progressive filling of the gap at the Fermi level with $Sn$-concentration, 
as one would expect. Experiments in
 Ge$_{1-x}$Sn$_{x}$ alloys \cite{atwater2000} have indicated that
the direct energy band gap decreases primarily through an increase
in alloy concentration (and applied coherency strain mainly  reduces the valence band DOS, 
instead of the magnitude of the gap).  

The results we present  were obtained adjusting the 
input lattice parameters at each alloy concentration using our respective Wien-2K obtained 
structural data (Table I). If, as it is usual in CPA, one uses the fixed lattice parameter 
corresponding to the perfect (x=0) lattice at all concentrations, one finds that the trends 
with concentration are much smaller, if observable. For instance, the resulting valence band obtained is almost unchanged by concentration increases of up to 25$\%$, 
thus not reflecting the effect of the difference of bandwidths (and shapes) 
of the two alloy components. 
In this sense we  have found that some improvement is gained by supplementing this effective field method  with information on the  structural changes undergone by the alloy as a function of concentration, considering that we found limitations of the traditional CPA description 
for Ge$_{1-x}$Sn$_{x}$ even at relatively low concentrations.
 
To better convey the information obtained by the FPLO+CPA approach, in Fig.~\ref{CPA-DOS-125}
we show the CPA results at x=0.125, namely the total density of states as well as the orbital-resolved partial densities of states. At this low $Sn$ concentration, the total DOS follows closely the Ge-host one.  
  
\begin{figure}[h !]
\includegraphics[angle=270, width=\columnwidth]{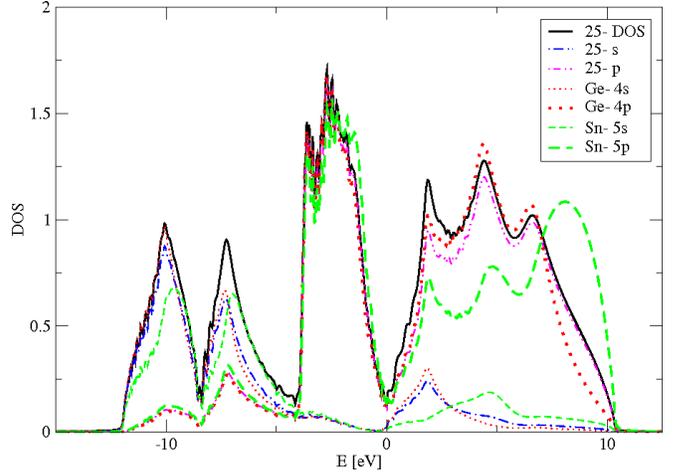}
\caption{(Color online) CPA results for Ge$_{0.75}$Sn$_{0.25}$, via FPLO-5:  
total density of states, orbital and orbital+species  resolved main DOS contributions, as functions 
of energy (references in plot).  }  
\label{CPA-DOS-25}
\end{figure}

Increasing concentration to x=0.25, in Fig.~\ref{CPA-DOS-25} we show the total DOS as well as the 
orbital-resolved partial DOS: here the most relevant ones correspond to $s$ orbitals (given by the 
$4s$ orbitals in Ge, and the $5s$ ones in $Sn$)  
 and $p$ orbitals ($4p$ of Ge, and $5p$ of $Sn$) while, for clarity, 
the relatively smaller $d$ components are not shown. 
Our results show good agreement with the valence band $s$ and $p$ partial DOS obtained by 
a previous LMTO Green function study.\cite{svane}
Basically, the spectral weight of $p$ orbitals strongly dominates around the Fermi level, 
while the $s$ orbitals are the most relevant at lower energies, for both species.

\begin{figure}[h !]
\includegraphics[angle=270, width=\columnwidth]{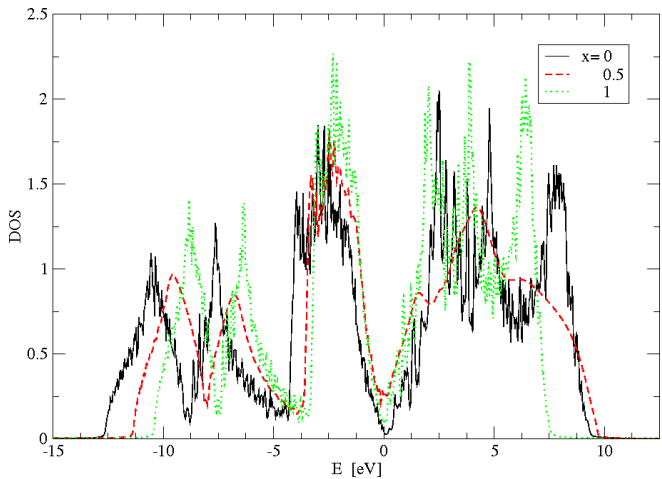}
\caption{(Color online) CPA    via    FPLO-5    calculated    DOS   for    $Ge$,    $Sn$
and Ge$_{0.5}$Sn$_{0.5}$.}
\label{CPA-DOS-III}
\end{figure}
Finally, in Fig.~\ref{CPA-DOS-III} we exhibit an example of the problems encountered by CPA,
based on the assumption that only substitutional $\alpha-Sn$ impurities are present,  
which  can not be overcome  by  taking  into  account  the lattice parameter changes 
as a  function of  $Sn$ concentration. Here the $x=1$ density of states is shown, in addition 
to the $x= 0, 0.5$ ones, showing that the smooth trend with $Sn$-concentration of 
Fig.~\ref{CPA-DOS-0to05},
consisting of the filling of the gap at the Fermi level, 
is not valid at larger concentrations ($x \ge 0.5$).
Contrary to expectations, the DOS at the Fermi level for pure $Sn$ is 
not the largest obtained for the alloy series. We believe that this  
provides a further indication of the presence of non-substitutional $Sn$ defects, 
and their relevance for a complete and consistent treatment of the problem.  
 
\section{Conclusion}
\label{conclusions}

We  have proposed a  mechanism  to understand  the peculiar
properties of  Ge$_{1-x}$Sn$_{x}$ alloys.  The main  assumption is the
existence of the $\beta-Sn$ defect,  occupying a divacancy in the $Ge$
host, which  imposes a severe strain  in the lattice  oppposite to the
one caused  by the $\alpha-Sn$ substitutional defect.   The other feature
of  the  $\beta$  defect  is  that it  causes  a  six-fold  octahedral
coordination on the  $Sn$ atom, favouring the nucleation  of white tin
inclusions which eventually segregate.

We  demonstrate  the plausibility  of  this  assumption by  performing
ab-initio electronic  calculations in supercells  containing different
local  defects. Starting  from the  single atom  $\alpha$  and $\beta$
defects,  and  increasing  the  number  of $Sn$  atoms,  in  order  to
estimate: the electronic energy,  the relaxed configuration around the
defects,  and the pressure caused  by them.   These studies
suggest  that the cubic  octahedral symmetry  is favoured  in clusters
containing few $Sn$ atoms, and  that the electronic energy per atom is
not  very  much  increased  by  accumulating  clusters  of  $Sn$.  The
calculated pressures  indicate that  the large positive  elastic field
around  these large  defects may  be  released by  the aggregation  of
$\beta$  defects,  which have  an  opposite  elastic  field.  We  have
observed  that building  up a  local pressure  in the  lattice  is not
necessarily correlated with the electronic energy of the cluster. This
is in agreement with  former ideas \cite{cardona, menendez2002} that there
are two  independent reasons to  account for the difficulties  to form
homogeneous Ge$_{1-x}$Sn$_{x}$ alloys, namely the electronic $d-$bands
which  make gray  $\alpha-Sn$  unstable at  room temperature  and the
large difference in the size of the atoms.  We have also observed that
the  energy of  two $\alpha$  defects increases  when they  are closer
together, also indicating that it  would be very difficult to obtain a
homogeneous  alloy  formed  only   by  $\alpha$  defects  since  their
mean-distance  would be  smaller than  a  unit cell.  Notice, that  if
strain  was  the  only  factor  jeopardizing  formation,  the  $x=0.5$
zincblende  compound  should  be  easy  to form,  since  it  would  be
strain-free: but this is not the case.

The knowledge  gained by the  local defect studies provided  the basic
ingredients to formulate a simple statistical model to investigate the
relative  concentration of  $\alpha$  and $\beta$  defects in  thermal
equilibrium as  a function of  the total $Sn$ concentration,  $x$. The
model focuses  on the dynamical evolution of  few local configurations
by  using   a  stochastic   matrix  containing  all   the  statistical
information to  calculate the transition between  one configuration to
another. The  results show  that,  at room  temperature,  there are  no
$\beta$ defects present for $x <  0.2 $. Thus, our model supports 
the experimental finding that, at concentrations below 20$\%$, 
$Sn$ only enters substitutionally with a local tetrahedral environment.  
 We find that
the number of $\beta$ defects  at a given concentration increases with
temperature.  We also found  that the  concentration at  which $\beta$
defects  start  appearing   decreases  at  larger  temperatures.  This
suggests  that one could  obtain homogeneous  alloys with  higher $Sn$
content if  one decreases  the temperature of  the thermal  bath. This
would  mean,  in  the  case  of  epitaxial growing  by  CVD  or  other
techniques, that  lowering the temperature  of the substrate  on which
the  alloy is  grown favours  the formation  of homogeneous  alloys at
higher  $Sn$  concentrations.    This  agrees  with  the  experimental
findings.\cite{menendez2002}  In   particular,  we predict that  $x=0.5$  homogeneous
alloys would be possible to form at temperatures below -90$^\circ$C.

In the last part of our work,  we have tried to gain more insight into
our problem  by using effective-field methods in  order to investigate
the  low concentration  region in  more detail.  There, we  found that
TB+VCA,    while    providing    a    good    description    of    the
substitutional-dominated regime,  fails to predict the  closing of the
electronic  gap at  $x=0.37$  seen in  experiments. Furthermore,  more
sophisticated mean-field techniques as  the CPA (even using a combined
ab-initio realistic bandstructure  with the single-site substitutional
CPA), does  not provide much improvement. However,  if one supplements
the     CPA     by     introducing    the     approximately     linear
concentration-dependence of  the lattice parameter,  obtained from our
ab-initio local defect calculations (in agreement with experiments), a
slight  improvement of  the  results is  obtained.  All this,  further
reinforces the  hypothesis put forward  in the present work.   We think
that  an extension  of the  CPA \cite{sces} able to  take into
account  more complex defects,  such as the non-substitutional $\beta$ defect, would certainly be extremely useful  to improve the  
description of the electronic properties of these fascinating and useful
alloys, and to study their effect on the large bowing of the direct band gap 
and the indirect to direct gap crossover.\cite{ferhat,menendez2006,SSC2006,harrison,PRB2008} 
Research along these lines is currently in progress.


\acknowledgments We thank M. Richter, K. Koepernik, H. Eschrig and 
the FPLO team, for the support and help to run their FPLO-CPA code, and for useful discussions; also  J. Menendez, P. Fulde and B. Bouhafs for useful references and discussions. We are specially grateful to Rub\'en Weht for his help in running additional FPLO-CPA cases and for his careful reading of the manuscript.
C.I.V. and J.D.F. are Investigadores Cient\'{\i}ficos of CONICET (Argentina). C.I.V. 
acknowledges support from CONICET and ANPCyT (PIP'5342 and PICT'38357 grants);  
from I.C.T.P., Trieste, Italy,  as {\it Regular Associate}; E.Peltzer and organizers 
of the 6th-FPLO Workshop (La Plata, Argentina, 2007),  and the hospitality  of Theoretical Physics
and  Worcester College, University of Oxford, U.K., where parts of this work were completed. 
R.A.B. thanks Centro At\'omico Bariloche and CMB Mathematical Institute,  University of Oxford, for hospitality; sabbatical grants from DGAPA (UNAM) and CONACyT are fully acknowledged.


\begin{thebibliography}{9}

\bibitem{temkin} R, J. Temkin, Solid State Commun. \textbf{11}, 1591 (1972).

\bibitem{goodman} C.H.L. Goodman, IEE Proc. I  \textbf{129}, 189 (1982).

\bibitem{oguz} S. Oguz, W. Paul, T.F. Deutsch, B.Y. Tsaur and D.V. Murphy, Appl. Phys. Lett. \textbf{43}, 848 (1983).

\bibitem{pearsall} T. P. Pearsall et al., Phys. Rev. Lett. \textbf{58}, 729 (1987).

\bibitem{cohen1986} Y.K. Vohra et al.,  Phys. Rev. Lett. \textbf{56}, 1944  (1986).

\bibitem{graytin} S. Grovs and W. Paul, Phys. Rev. Lett. \textbf{11}, 194 (1963).


\bibitem{jenkins} D. W. Jenkins and John D. Dow,  Phys. Rev. B \textbf{36}, 7994 (1987).

\bibitem{baldereschi} K. A. Maeder and A. Baldereschi, Solid State Commun. \textbf{69}, 1123 (1989).

\bibitem{atwater} G. He and H.A. Atwater, Phys. Rev. Lett. \textbf{79}, 1937 (1997).

\bibitem{menendez2002} M. Bauer et al., Applied Phys. Lett. \textbf{81}, 2992 (2002).

\bibitem{SSC2003}  M. R. Bauer et al., Solid State Commun. \textbf{127}, 355 (2003).

\bibitem{ladron} H. P\'erez Ladr\'on de Guevara, A.G. Rodriguez, H. Navarro-Contreras and M.A. Vidal, Applied Phys. Lett. \textbf{84}, 4532 (2004).

\bibitem{atwater2006} R. Ragan, J.E. Guyer, E. Meserole, M.S. Goorski and H.A. Atwater,  Phys. Rev. B \textbf{73}, 235303 (2006).

\bibitem{cohen2007} J.D. Sau and M.L. Cohen,  Phys. Rev. B \textbf{75}, 45208 (2007).


\bibitem{APL2004} S. F. Li, M.R. Bauer, J. Menendez and J. Kouvetakis, Applied Phys. Lett. \textbf{84}, 867 (2004).

\bibitem{menendez2006} V.R. d' Costa et al.,  Phys. Rev. B \textbf{73}, 125207 (2006).

\bibitem{barrio-am} I. Chambouleyron, F. Marques, P.H. Dionisio, I.J.R. Baumvol and R.A. Barrio,  J. Appl. Phys. \textbf{66}, 2083 (1989).

\bibitem{mrs} P. Boolchand and C.C. Koch,  J. Mater. Res. \textbf{7}, 2876 (1992).

\bibitem{VCA} L. Nordheim, Ann. Phys. (Lipz) \textbf{9},
  607 (1931).

\bibitem{soven} P. Soven, Phys. Rev.  \textbf{156}, 809 (1967).

\bibitem{ferhat} Y. Chibane, B. Bouhafs and M. Ferhat, Phys. Stat. Sol. $(b)$  \textbf{240}, 116 (2003).

\bibitem{j1}P. Blaha, K. Schwarz, G.K.H. Madsen, D. Kvasnicka, J. Luitz,
\textit{``WIEN2k, An Augmented Plane Wave + Local Orbitals Program for
Calculating Crystal Properties''}, (Karlheinz Schwarz,
T.U. Wien, Austria, 2001): ISBN 3-9501031-1-2.

\bibitem{j2} K. Schwarz, P. Blaha, and G. K. H. Madsen, 
Comput. Phys. Commun. \textbf{147}, 71 (2002).

\bibitem{j3} K. Schwarz and P. Blaha, Comput. Mater. Sci. \textbf{28}, 259 (2003).

\bibitem{j4} J.P. Perdew, S. Burke, M. Ernzerhof, Phys. Lett. \textbf{77}, 3865 (1996).

\bibitem{j5} H.J. Monkhorst and J.D. Pack, Phys. Rev. B \textbf{13}, 5188 (1976).

\bibitem{paperRaf1-stochastic}  R. A. Barrio, R. Kerner, M. Micoulaut 
and G. G. Naumis, J. Phys.: Cond. Matter \textbf{9}, 9219 (1997). 

\bibitem{book-Kerner} R. Kerner,  \textit{"Models of Agglomeration and Glass Transition''}, 
 Imperial College Press, London (2006).

\bibitem{cardona} T. Brudevoll, D.S. Citrin, N.E. Christensen and M. Cardona, Phys. Rev. B \textbf{48}, 17128 (1993).

\bibitem{cpa-semic} M. Jaros,  Rep. Prog. Phys.  \textbf{48}, 1091 (1985).

\bibitem{cpa-gesi} D. Stroud and H. Ehrenreich, Phys. Rev. B \textbf{2}, 3197
  (1970); S. Krishnamurthy, A. Sher and A. B. Chen,  Phys. Rev. B \textbf{33}, 1026 (1986).

\bibitem{harrison} P. Moontragoon, Z. Ikonic and P.Harrison, Semicond. Sci. Technol.  \textbf{22}, 742 (2007).

\bibitem{FPLO} K. Koepernik and H. Eschrig, Phys. Rev. B\textbf{ 59}, 1743 (1999);
I. Opahle, K. Koepernik, and H. Eschrig, Phys. Rev. B \textbf{ 60},
               14035 (1999);

\bibitem{FPLO-CPA} K. Koepernik, B. Velicky, R. Hayn, and H. Eschrig, Phys. Rev. B
  \textbf{ 55}, 5717 (1997).

\bibitem{BEB} J. Blackman, D. Esterling and N. Berk, Phys. Rev. B \textbf{4},
  2412 (1971); for a review of approaches to disorder see
  e.g.  A. Gonis, \textit{"Green functions for ordered and disordered systems''}, 
  North-Holland, Amsterdam (1992).  


\bibitem{atwater2000} R. Ragan and H.A. Atwater, Applied
  Phys. Lett. \textbf{77}, 3418 (2000).

\bibitem{svane} A. Svane,  J. Phys. C: Solid State Phys. \textbf{21},
  5369 (1988).

 \bibitem{sces} C.I.  Ventura and R. A. Barrio, Physica  B 281-282, 855 (2000).

\bibitem{SSC2006} V. R. d' Costa et al.,  Solid State Commun. \textbf{138},
  309 (2006).

\bibitem{PRB2008} K. Alberi et al., Phys. Rev. B  \textbf{ 77}, 73202 (2008).

\end{thebibliography}
\end{document}